# A GENERAL EXPRESSION FOR THE 14$^{th}$ CHERN FORM

C. C. Briggs
*Center for Academic Computing, Penn State University, University Park, PA 16802*
Friday, April 16, 1999

**Abstract.** A general expression is given for the 14$^{th}$ Chern form in terms of simple polynomial concomitants of the curvature 2-form for $n$-dimensional differentiable manifolds having a general linear connection.

This letter presents a general expression for the 14$^{th}$ Chern form in terms of simple polynomial concomitants of the curvature 2-form for $n$-dimensional differentiable manifolds having a general linear connection.

The $p^{th}$ Chern forms[1] $c_{(p)}$ representing the corresponding $p^{th}$ Chern classes of such a manifold $M$ can be defined by[2,3]

$$c_{(p)} \equiv \begin{cases} 1, & \text{if } p = 0 \\ \dfrac{i^p}{2^p \pi^p} \Omega_{[i_1}{}^{i_1} \wedge \Omega_{i_2}{}^{i_2} \wedge \ldots \wedge \Omega_{i_p]}{}^{i_p}, & \text{if } p > 0 \end{cases}, \quad (1)$$

where $\Omega_a{}^b$ is the curvature 2-form of $M$.

Some numerical properties of $c_{(p)}$ for $p = 14$ appear in Table 1. A general expression for $c_{(p)}$ for $p = 14$ appears in Eq. (3) in terms of the comcomitants $\operatorname{tr}^A(\Omega^B)$ defined by

$$\operatorname{tr}^A(\Omega^B) \equiv \Omega_{i_2}{}^{i_1} \wedge \Omega_{i_3}{}^{i_2} \wedge \ldots \wedge \Omega_{i_1}{}^{i_B} \wedge \Omega_{i_{B+2}}{}^{i_{B+1}} \wedge \Omega_{i_{B+3}}{}^{i_{B+2}} \wedge \ldots \wedge \Omega_{i_{B+1}}{}^{i_{2B}} \wedge \ldots \wedge \Omega_{i_{(A-1)B+2}}{}^{i_{(A-1)B+1}} \wedge \Omega_{i_{(A-1)B+3}}{}^{i_{(A-1)B+2}} \wedge \ldots \wedge \Omega_{i_{(A-1)B+1}}{}^{i_{AB}}, \quad (2)$$

where $A$ and $B$ are integers $\geq 1$.

### TABLE 1. SOME NUMERICAL PROPERTIES OF $c_{(p)}$ FOR $p = 14$

| QUANTITY | ORDER | CURVATURE DEPENDENCE | MINIMUM NUMBER OF DIMENSIONS | NUMBER OF TERMS | 1$^{st}$ OVERALL NUMERICAL FACTOR | 2$^{nd}$ OVERALL NUMERICAL FACTOR | NUMBER OF PERMUTATIONS COMPREHENDED |
|---|---|---|---|---|---|---|---|
| $c_{(p)}$ | $p$ | — | $2p$ | — | $\dfrac{i^p}{2^p \pi^p}$ | $\dfrac{i^p}{2^p \pi^p p!}$ | $p!$ |
| $c_{(14)}$ | 14 | Quattuordecic | 28 | **135** | $-\dfrac{1}{16{,}384\,\pi^{14}}$ | $-\dfrac{1}{1{,}428{,}329{,}123{,}020{,}800\,\pi^{14}}$ | 87,178,291,200 |

## 14$^{th}$ CHERN FORM

$$c_{(14)} = \frac{i^{14}}{2^{14}\pi^{14}} \Omega_{[i_1}{}^{i_1} \wedge \Omega_{i_2}{}^{i_2} \wedge \Omega_{i_3}{}^{i_3} \wedge \Omega_{i_4}{}^{i_4} \wedge \Omega_{i_5}{}^{i_5} \wedge \Omega_{i_6}{}^{i_6} \wedge \Omega_{i_7}{}^{i_7} \wedge \Omega_{i_8}{}^{i_8} \wedge \Omega_{i_9}{}^{i_9} \wedge \Omega_{i_{10}}{}^{i_{10}} \wedge \Omega_{i_{11}}{}^{i_{11}} \wedge \Omega_{i_{12}}{}^{i_{12}} \wedge \Omega_{i_{13}}{}^{i_{13}} \wedge \Omega_{i_{14}]}{}^{i_{14}} \quad (3)$$

$$= \frac{i^{14}}{2^{14}\pi^{14}\,14!} \Big( -6{,}227{,}020{,}800\,\operatorname{tr}(\Omega^{14}) + 6{,}706{,}022{,}400\,\operatorname{tr}(\Omega^{13}) \wedge \operatorname{tr}(\Omega) + 3{,}632{,}428{,}800\,\operatorname{tr}(\Omega^{12}) \wedge \operatorname{tr}(\Omega^2) - 3{,}632{,}428{,}800\,\operatorname{tr}(\Omega^{12}) \wedge \operatorname{tr}^2(\Omega) +$$

$$+ 2{,}641{,}766{,}400\,\operatorname{tr}(\Omega^{11}) \wedge \operatorname{tr}(\Omega^3) - 3{,}962{,}649{,}600\,\operatorname{tr}(\Omega^{11}) \wedge \operatorname{tr}(\Omega^2) \wedge \operatorname{tr}(\Omega) + 1{,}320{,}883{,}200\,\operatorname{tr}(\Omega^{11}) \wedge \operatorname{tr}^3(\Omega) +$$

$$+ 2{,}179{,}457{,}280\,\operatorname{tr}(\Omega^{10}) \wedge \operatorname{tr}(\Omega^4) - 2{,}905{,}943{,}040\,\operatorname{tr}(\Omega^{10}) \wedge \operatorname{tr}(\Omega^3) \wedge \operatorname{tr}(\Omega) - 1{,}089{,}728{,}640\,\operatorname{tr}(\Omega^{10}) \wedge \operatorname{tr}^2(\Omega^2) +$$

$$+ 2{,}179{,}457{,}280\,\operatorname{tr}(\Omega^{10}) \wedge \operatorname{tr}(\Omega^2) \wedge \operatorname{tr}^2(\Omega) - 363{,}242{,}880\,\operatorname{tr}(\Omega^{10}) \wedge \operatorname{tr}^4(\Omega) + 1{,}937{,}295{,}360\,\operatorname{tr}(\Omega^9) \wedge \operatorname{tr}(\Omega^5) -$$

$$- 2{,}421{,}619{,}200\,\operatorname{tr}(\Omega^9) \wedge \operatorname{tr}(\Omega^4) \wedge \operatorname{tr}(\Omega) - 1{,}614{,}412{,}800\,\operatorname{tr}(\Omega^9) \wedge \operatorname{tr}(\Omega^3) \wedge \operatorname{tr}(\Omega^2) + 1{,}614{,}412{,}800\,\operatorname{tr}(\Omega^9) \wedge \operatorname{tr}(\Omega^3) \wedge \operatorname{tr}^2(\Omega) +$$

$$+ 1{,}210{,}809{,}600\,\operatorname{tr}(\Omega^9) \wedge \operatorname{tr}^2(\Omega^2) \wedge \operatorname{tr}(\Omega) - 807{,}206{,}400\,\operatorname{tr}(\Omega^9) \wedge \operatorname{tr}(\Omega^2) \wedge \operatorname{tr}^3(\Omega) + 80{,}720{,}640\,\operatorname{tr}(\Omega^9) \wedge \operatorname{tr}^5(\Omega) +$$

$$+ 1{,}816{,}214{,}400\,\operatorname{tr}(\Omega^8) \wedge \operatorname{tr}(\Omega^6) - 2{,}179{,}457{,}280\,\operatorname{tr}(\Omega^8) \wedge \operatorname{tr}(\Omega^5) \wedge \operatorname{tr}(\Omega) - 1{,}362{,}160{,}800\,\operatorname{tr}(\Omega^8) \wedge \operatorname{tr}(\Omega^4) \wedge \operatorname{tr}(\Omega^2) +$$

$$+ 1{,}362{,}160{,}800\,\operatorname{tr}(\Omega^8) \wedge \operatorname{tr}(\Omega^4) \wedge \operatorname{tr}^2(\Omega) - 605{,}404{,}800\,\operatorname{tr}(\Omega^8) \wedge \operatorname{tr}^2(\Omega^3) + 1{,}816{,}214{,}400\,\operatorname{tr}(\Omega^8) \wedge \operatorname{tr}(\Omega^3) \wedge \operatorname{tr}(\Omega^2) \wedge \operatorname{tr}(\Omega) -$$

$$- 605{,}404{,}800\,\operatorname{tr}(\Omega^8) \wedge \operatorname{tr}(\Omega^3) \wedge \operatorname{tr}^3(\Omega) + 227{,}026{,}800\,\operatorname{tr}(\Omega^8) \wedge \operatorname{tr}^3(\Omega^2) - 681{,}080{,}400\,\operatorname{tr}(\Omega^8) \wedge \operatorname{tr}^2(\Omega^2) \wedge \operatorname{tr}^2(\Omega) +$$

$$+ 227{,}026{,}800\,\operatorname{tr}(\Omega^8) \wedge \operatorname{tr}(\Omega^2) \wedge \operatorname{tr}^4(\Omega) - 15{,}135{,}120\,\operatorname{tr}(\Omega^8) \wedge \operatorname{tr}^6(\Omega) + 889{,}574{,}400\,\operatorname{tr}^2(\Omega^7) - 2{,}075{,}673{,}600\,\operatorname{tr}(\Omega^7) \wedge \operatorname{tr}(\Omega^6) \wedge \operatorname{tr}(\Omega) -$$

$$- 1{,}245{,}404{,}160\,\operatorname{tr}(\Omega^7) \wedge \operatorname{tr}(\Omega^5) \wedge \operatorname{tr}(\Omega^2) + 1{,}245{,}404{,}160\,\operatorname{tr}(\Omega^7) \wedge \operatorname{tr}(\Omega^5) \wedge \operatorname{tr}^2(\Omega) - 1{,}037{,}836{,}800\,\operatorname{tr}(\Omega^7) \wedge \operatorname{tr}(\Omega^4) \wedge \operatorname{tr}(\Omega^3) +$$

$$+ 1{,}556{,}755{,}200\,\operatorname{tr}(\Omega^7) \wedge \operatorname{tr}(\Omega^4) \wedge \operatorname{tr}(\Omega^2) \wedge \operatorname{tr}(\Omega) - 518{,}918{,}400\,\operatorname{tr}(\Omega^7) \wedge \operatorname{tr}(\Omega^4) \wedge \operatorname{tr}^3(\Omega) + 691{,}891{,}200\,\operatorname{tr}(\Omega^7) \wedge \operatorname{tr}^2(\Omega^3) \wedge \operatorname{tr}(\Omega) +$$

$$+ 518{,}918{,}400\,\operatorname{tr}(\Omega^7) \wedge \operatorname{tr}(\Omega^3) \wedge \operatorname{tr}^2(\Omega^2) - 1{,}037{,}836{,}800\,\operatorname{tr}(\Omega^7) \wedge \operatorname{tr}(\Omega^3) \wedge \operatorname{tr}(\Omega^2) \wedge \operatorname{tr}^2(\Omega) + 172{,}972{,}800\,\operatorname{tr}(\Omega^7) \wedge \operatorname{tr}(\Omega^3) \wedge \operatorname{tr}^4(\Omega) -$$

$$- 259{,}459{,}200\,\operatorname{tr}(\Omega^7) \wedge \operatorname{tr}^3(\Omega^2) \wedge \operatorname{tr}(\Omega) + 259{,}459{,}200\,\operatorname{tr}(\Omega^7) \wedge \operatorname{tr}^2(\Omega^2) \wedge \operatorname{tr}^3(\Omega) - 51{,}891{,}840\,\operatorname{tr}(\Omega^7) \wedge \operatorname{tr}(\Omega^2) \wedge \operatorname{tr}^5(\Omega) +$$

$$+ 2{,}471{,}040\,\operatorname{tr}(\Omega^7) \wedge \operatorname{tr}^7(\Omega) - 605{,}404{,}800\,\operatorname{tr}^2(\Omega^6) \wedge \operatorname{tr}(\Omega^2) + 605{,}404{,}800\,\operatorname{tr}^2(\Omega^6) \wedge \operatorname{tr}^2(\Omega) - 968{,}647{,}680\,\operatorname{tr}(\Omega^6) \wedge \operatorname{tr}(\Omega^5) \wedge \operatorname{tr}(\Omega^3) +$$

$$+ 1{,}452{,}971{,}520\,\operatorname{tr}(\Omega^6) \wedge \operatorname{tr}(\Omega^5) \wedge \operatorname{tr}(\Omega^2) \wedge \operatorname{tr}(\Omega) - 484{,}323{,}840\,\operatorname{tr}(\Omega^6) \wedge \operatorname{tr}(\Omega^5) \wedge \operatorname{tr}^3(\Omega) - 454{,}053{,}600\,\operatorname{tr}(\Omega^6) \wedge \operatorname{tr}^2(\Omega^4) +$$

$$+ 1{,}210{,}809{,}600\,\operatorname{tr}(\Omega^6) \wedge \operatorname{tr}(\Omega^4) \wedge \operatorname{tr}(\Omega^3) \wedge \operatorname{tr}(\Omega) + 454{,}053{,}600\,\operatorname{tr}(\Omega^6) \wedge \operatorname{tr}(\Omega^4) \wedge \operatorname{tr}^2(\Omega^2) - 908{,}107{,}200\,\operatorname{tr}(\Omega^6) \wedge \operatorname{tr}(\Omega^4) \wedge \operatorname{tr}(\Omega^2) \wedge \operatorname{tr}^2(\Omega) +$$

$$+ 151{,}351{,}200\,\operatorname{tr}(\Omega^6) \wedge \operatorname{tr}(\Omega^4) \wedge \operatorname{tr}^4(\Omega) + 403{,}603{,}200\,\operatorname{tr}(\Omega^6) \wedge \operatorname{tr}^2(\Omega^3) \wedge \operatorname{tr}(\Omega^2) - 403{,}603{,}200\,\operatorname{tr}(\Omega^6) \wedge \operatorname{tr}^2(\Omega^3) \wedge \operatorname{tr}^2(\Omega) -$$

$$- 605{,}404{,}800\,\operatorname{tr}(\Omega^6) \wedge \operatorname{tr}(\Omega^3) \wedge \operatorname{tr}^2(\Omega^2) \wedge \operatorname{tr}(\Omega) + 403{,}603{,}200\,\operatorname{tr}(\Omega^6) \wedge \operatorname{tr}(\Omega^3) \wedge \operatorname{tr}(\Omega^2) \wedge \operatorname{tr}^3(\Omega) - 40{,}360{,}320\,\operatorname{tr}(\Omega^6) \wedge \operatorname{tr}(\Omega^3) \wedge \operatorname{tr}^5(\Omega) -$$

$$- 37{,}837{,}800\,\operatorname{tr}(\Omega^6) \wedge \operatorname{tr}^4(\Omega^2) + 151{,}351{,}200\,\operatorname{tr}(\Omega^6) \wedge \operatorname{tr}^3(\Omega^2) \wedge \operatorname{tr}^2(\Omega) - 75{,}675{,}600\,\operatorname{tr}(\Omega^6) \wedge \operatorname{tr}^2(\Omega^2) \wedge \operatorname{tr}^4(\Omega) +$$

$$+ 10{,}090{,}080\,\operatorname{tr}(\Omega^6) \wedge \operatorname{tr}(\Omega^2) \wedge \operatorname{tr}^6(\Omega) - 360{,}360\,\operatorname{tr}(\Omega^6) \wedge \operatorname{tr}^8(\Omega) - 435{,}891{,}456\,\operatorname{tr}^2(\Omega^5) \wedge \operatorname{tr}(\Omega^4) + 581{,}188{,}608\,\operatorname{tr}^2(\Omega^5) \wedge \operatorname{tr}(\Omega^3) \wedge \operatorname{tr}(\Omega) +$$

$$+ 217{,}945{,}728\,\operatorname{tr}^2(\Omega^5) \wedge \operatorname{tr}^2(\Omega^2) - 435{,}891{,}456\,\operatorname{tr}^2(\Omega^5) \wedge \operatorname{tr}(\Omega^2) \wedge \operatorname{tr}^2(\Omega) + 72{,}648{,}576\,\operatorname{tr}^2(\Omega^5) \wedge \operatorname{tr}^4(\Omega) +$$

$$+ 544{,}864{,}320\,\operatorname{tr}(\Omega^5) \wedge \operatorname{tr}^2(\Omega^4) \wedge \operatorname{tr}(\Omega) + 726{,}485{,}760\,\operatorname{tr}(\Omega^5) \wedge \operatorname{tr}(\Omega^4) \wedge \operatorname{tr}(\Omega^3) \wedge \operatorname{tr}(\Omega^2) - 726{,}485{,}760\,\operatorname{tr}(\Omega^5) \wedge \operatorname{tr}(\Omega^4) \wedge \operatorname{tr}(\Omega^3) \wedge \operatorname{tr}^2(\Omega) -$$

$$- 544{,}864{,}320\,\operatorname{tr}(\Omega^5) \wedge \operatorname{tr}(\Omega^4) \wedge \operatorname{tr}^2(\Omega^2) \wedge \operatorname{tr}(\Omega) + 363{,}242{,}880\,\operatorname{tr}(\Omega^5) \wedge \operatorname{tr}(\Omega^4) \wedge \operatorname{tr}(\Omega^2) \wedge \operatorname{tr}^3(\Omega) - 36{,}324{,}288\,\operatorname{tr}(\Omega^5) \wedge \operatorname{tr}(\Omega^4) \wedge \operatorname{tr}^5(\Omega) +$$

---

$+ 107{,}627{,}520\ \mathrm{tr}(\Omega^5) \wedge \mathrm{tr}^3(\Omega^3) - 484{,}323{,}840\ \mathrm{tr}(\Omega^5) \wedge \mathrm{tr}^2(\Omega^3) \wedge \mathrm{tr}(\Omega^2) \wedge \mathrm{tr}(\Omega) + 161{,}441{,}280\ \mathrm{tr}(\Omega^5) \wedge \mathrm{tr}^2(\Omega^3) \wedge \mathrm{tr}^3(\Omega) -$

$- 121{,}080{,}960\ \mathrm{tr}(\Omega^5) \wedge \mathrm{tr}(\Omega^3) \wedge \mathrm{tr}^3(\Omega^2) + 363{,}242{,}880\ \mathrm{tr}(\Omega^5) \wedge \mathrm{tr}(\Omega^3) \wedge \mathrm{tr}^2(\Omega^2) \wedge \mathrm{tr}^2(\Omega) - 121{,}080{,}960\ \mathrm{tr}(\Omega^5) \wedge \mathrm{tr}(\Omega^3) \wedge \mathrm{tr}(\Omega^2) \wedge \mathrm{tr}^4(\Omega) +$

$+ 8{,}072{,}064\ \mathrm{tr}(\Omega^5) \wedge \mathrm{tr}(\Omega^3) \wedge \mathrm{tr}^6(\Omega) + 45{,}405{,}360\ \mathrm{tr}(\Omega^5) \wedge \mathrm{tr}^4(\Omega^2) \wedge \mathrm{tr}(\Omega) - 60{,}540{,}480\ \mathrm{tr}(\Omega^5) \wedge \mathrm{tr}^3(\Omega^2) \wedge \mathrm{tr}^3(\Omega) +$

$+ 18{,}162{,}144\ \mathrm{tr}(\Omega^5) \wedge \mathrm{tr}^2(\Omega^2) \wedge \mathrm{tr}^5(\Omega) - 1{,}729{,}728\ \mathrm{tr}(\Omega^5) \wedge \mathrm{tr}(\Omega^2) \wedge \mathrm{tr}^7(\Omega) + 48{,}048\ \mathrm{tr}(\Omega^5) \wedge \mathrm{tr}^9(\Omega) + 113{,}513{,}400\ \mathrm{tr}^3(\Omega^4) \wedge \mathrm{tr}(\Omega^2) -$

$- 113{,}513{,}400\ \mathrm{tr}^3(\Omega^4) \wedge \mathrm{tr}^2(\Omega) + 151{,}351{,}200\ \mathrm{tr}^2(\Omega^4) \wedge \mathrm{tr}^2(\Omega^3) - 454{,}053{,}600\ \mathrm{tr}^2(\Omega^4) \wedge \mathrm{tr}(\Omega^3) \wedge \mathrm{tr}(\Omega^2) \wedge \mathrm{tr}(\Omega) +$

$+ 151{,}351{,}200\ \mathrm{tr}^2(\Omega^4) \wedge \mathrm{tr}(\Omega^3) \wedge \mathrm{tr}^3(\Omega) - 56{,}756{,}700\ \mathrm{tr}^2(\Omega^4) \wedge \mathrm{tr}^3(\Omega^2) + 170{,}270{,}100\ \mathrm{tr}^2(\Omega^4) \wedge \mathrm{tr}^2(\Omega^2) \wedge \mathrm{tr}^2(\Omega) -$

$- 56{,}756{,}700\ \mathrm{tr}^2(\Omega^4) \wedge \mathrm{tr}(\Omega^2) \wedge \mathrm{tr}^4(\Omega) + 3{,}783{,}780\ \mathrm{tr}^2(\Omega^4) \wedge \mathrm{tr}^6(\Omega) - 134{,}534{,}400\ \mathrm{tr}(\Omega^4) \wedge \mathrm{tr}^3(\Omega^3) \wedge \mathrm{tr}(\Omega) -$

$- 151{,}351{,}200\ \mathrm{tr}(\Omega^4) \wedge \mathrm{tr}^2(\Omega^3) \wedge \mathrm{tr}^2(\Omega^2) + 302{,}702{,}400\ \mathrm{tr}(\Omega^4) \wedge \mathrm{tr}^2(\Omega^3) \wedge \mathrm{tr}(\Omega^2) \wedge \mathrm{tr}^2(\Omega) - 50{,}450{,}400\ \mathrm{tr}(\Omega^4) \wedge \mathrm{tr}^2(\Omega^3) \wedge \mathrm{tr}^4(\Omega) +$

$+ 151{,}351{,}200\ \mathrm{tr}(\Omega^4) \wedge \mathrm{tr}(\Omega^3) \wedge \mathrm{tr}^3(\Omega^2) \wedge \mathrm{tr}(\Omega) - 151{,}351{,}200\ \mathrm{tr}(\Omega^4) \wedge \mathrm{tr}(\Omega^3) \wedge \mathrm{tr}^2(\Omega^2) \wedge \mathrm{tr}^3(\Omega) +$

$+ 30{,}270{,}240\ \mathrm{tr}(\Omega^4) \wedge \mathrm{tr}(\Omega^3) \wedge \mathrm{tr}(\Omega^2) \wedge \mathrm{tr}^5(\Omega) - 1{,}441{,}440\ \mathrm{tr}(\Omega^4) \wedge \mathrm{tr}(\Omega^3) \wedge \mathrm{tr}^7(\Omega) + 5{,}675{,}670\ \mathrm{tr}(\Omega^4) \wedge \mathrm{tr}^5(\Omega^2) -$

$- 28{,}378{,}350\ \mathrm{tr}(\Omega^4) \wedge \mathrm{tr}^4(\Omega^2) \wedge \mathrm{tr}^2(\Omega) + 18{,}918{,}900\ \mathrm{tr}(\Omega^4) \wedge \mathrm{tr}^3(\Omega^2) \wedge \mathrm{tr}^4(\Omega) - 3{,}783{,}780\ \mathrm{tr}(\Omega^4) \wedge \mathrm{tr}^2(\Omega^2) \wedge \mathrm{tr}^6(\Omega) +$

$+ 270{,}270\ \mathrm{tr}(\Omega^4) \wedge \mathrm{tr}(\Omega^2) \wedge \mathrm{tr}^8(\Omega) - 6{,}006\ \mathrm{tr}(\Omega^4) \wedge \mathrm{tr}^{10}(\Omega) - 22{,}422{,}400\ \mathrm{tr}^4(\Omega^3) \wedge \mathrm{tr}(\Omega^2) + 22{,}422{,}400\ \mathrm{tr}^4(\Omega^3) \wedge \mathrm{tr}^2(\Omega) +$

$+ 67{,}267{,}200\ \mathrm{tr}^3(\Omega^3) \wedge \mathrm{tr}^2(\Omega^2) \wedge \mathrm{tr}(\Omega) - 44{,}844{,}800\ \mathrm{tr}^3(\Omega^3) \wedge \mathrm{tr}(\Omega^2) \wedge \mathrm{tr}^3(\Omega) + 4{,}484{,}480\ \mathrm{tr}^3(\Omega^3) \wedge \mathrm{tr}^5(\Omega) +$

$+ 12{,}612{,}600\ \mathrm{tr}^2(\Omega^3) \wedge \mathrm{tr}^4(\Omega^2) - 50{,}450{,}400\ \mathrm{tr}^2(\Omega^3) \wedge \mathrm{tr}^3(\Omega^2) \wedge \mathrm{tr}^2(\Omega) + 25{,}225{,}200\ \mathrm{tr}^2(\Omega^3) \wedge \mathrm{tr}^2(\Omega^2) \wedge \mathrm{tr}^4(\Omega) -$

$- 3{,}363{,}360\ \mathrm{tr}^2(\Omega^3) \wedge \mathrm{tr}(\Omega^2) \wedge \mathrm{tr}^6(\Omega) + 120{,}120\ \mathrm{tr}^2(\Omega^3) \wedge \mathrm{tr}^8(\Omega) - 7{,}567{,}560\ \mathrm{tr}(\Omega^3) \wedge \mathrm{tr}^5(\Omega^2) \wedge \mathrm{tr}(\Omega) +$

$+ 12{,}612{,}600\ \mathrm{tr}(\Omega^3) \wedge \mathrm{tr}^4(\Omega^2) \wedge \mathrm{tr}^3(\Omega) - 5{,}045{,}040\ \mathrm{tr}(\Omega^3) \wedge \mathrm{tr}^3(\Omega^2) \wedge \mathrm{tr}^5(\Omega) + 720{,}720\ \mathrm{tr}(\Omega^3) \wedge \mathrm{tr}^2(\Omega^2) \wedge \mathrm{tr}^7(\Omega) -$

$- 40{,}040\ \mathrm{tr}(\Omega^3) \wedge \mathrm{tr}(\Omega^2) \wedge \mathrm{tr}^9(\Omega) + 728\ \mathrm{tr}(\Omega^3) \wedge \mathrm{tr}^{11}(\Omega) - 135{,}135\ \mathrm{tr}^7(\Omega^2) + 945{,}945\ \mathrm{tr}^6(\Omega^2) \wedge \mathrm{tr}^2(\Omega) - 945{,}945\ \mathrm{tr}^5(\Omega^2) \wedge \mathrm{tr}^4(\Omega) +$

$+ 315{,}315\ \mathrm{tr}^4(\Omega^2) \wedge \mathrm{tr}^6(\Omega) - 45{,}045\ \mathrm{tr}^3(\Omega^2) \wedge \mathrm{tr}^8(\Omega) + 3{,}003\ \mathrm{tr}^2(\Omega^2) \wedge \mathrm{tr}^{10}(\Omega) - 91\ \mathrm{tr}(\Omega^2) \wedge \mathrm{tr}^{12}(\Omega) + \mathrm{tr}^{14}(\Omega))$

## CONCLUDING REMARK

For a check, note that the magnitudes of the numerical factors in the preceding expression for $c_{(14)}$ in Eq. (3) add up—aside from the overall numerical factor—to the number $14! = 87{,}178{,}291{,}200$ of covariant index permutations comprehended by $c_{(14)}$ per Eq. (1).